\documentclass[conference,letterpaper]{IEEEtran}
\usepackage[T1]{fontenc}
\ifdefined\pdfminorversion
\fi
\usepackage{amsmath,amssymb,bm,graphicx,booktabs,cite,xcolor,url,placeins,balance}
\usepackage[hidelinks]{hyperref}
\newcommand{\mat}[1]{\bm{#1}}
\newcommand{\Id}{\mat I_2}
\newcommand{\Gdel}{\mat G_{\mathrm{del},m}}
\newcommand{\Gi}{\mat G_{iL}}
\newcommand{\Gim}{\mat G_{iL,\mathrm{MFC}}}
\newcommand{\Ys}{\mat Y_s}
\newcommand{\Co}{\mat C_o}
\newcommand{\Ym}{\mat Y_{s,\mathrm{MFC}}}

\graphicspath{{figures/}}

\title{Model-Free Output-Admittance Reshaping for\\
Stability Enhancement of Grid-Connected Inverters}
\author{\IEEEauthorblockN{Runnan Wang\IEEEauthorrefmark{1},
Wanrong Li\IEEEauthorrefmark{1},
Huawei Yuan\IEEEauthorrefmark{2}, and
Sinan Li\IEEEauthorrefmark{1}}
\IEEEauthorblockA{\IEEEauthorrefmark{1}School of Electrical and Computer Engineering,
The University of Sydney, Sydney, NSW, Australia\\
Emails: runnan.wang@sydney.edu.au; wanrong.li@sydney.edu.au; sinan.li@sydney.edu.au}
\IEEEauthorblockA{\IEEEauthorrefmark{2}Department of Electrical Engineering,
City University of Hong Kong, Hong Kong SAR, China\\
Email: huawyuan@cityu.edu.hk}}
\begin{document}
\maketitle

\begin{abstract}
Grid-connected inverters can be unstable when their current-control loops interact with an uncertain external network. Retuning the existing controller to improve damping also alters current tracking because tracking dynamics and output admittance are coupled. This paper adds a local model-free control (MFC) loop around the existing current controller to reduce output admittance while closely preserving nominal current-tracking dynamics. The added loop uses local inverter parameters without an external-grid or neighboring-converter model. A Norton formulation separates reference tracking from the current response to terminal-voltage disturbances and establishes a tracking-preservation condition for admittance reshaping. Reducing the disturbance response makes the inverter behave more like an ideal controlled current source. Averaged-model simulations show reductions exceeding 65\% in the magnitudes of all four $dq$ output-admittance elements over 10--200~Hz, with closely preserved current-tracking dynamics. Generalized Nyquist analysis predicts small-signal stability for all sampled grid-impedance cases with MFC, compared with 90.0\% without MFC.
\end{abstract}
\begin{IEEEkeywords}
Current control, grid-connected inverters, impedance-based stability, model-free control, output admittance.
\end{IEEEkeywords}

\section{Introduction}
Power-electronic interfaces in renewable generation and storage can become unstable through converter--network interactions despite stable local current loops \cite{sun2011,wen2017}. Network impedance varies with configuration, while neighboring converters introduce operating-point-dependent dynamics. Since detailed external-system models may be unavailable, stability enhancement should require limited information about the surrounding network.

For a current-controlled inverter, output admittance quantifies the terminal-voltage-to-current response. Reducing it makes the inverter closer to an ideal controlled current source and weakens its interaction with the network. Current-feedback active damping suppresses filter resonance and can be interpreted through virtual impedance \cite{wang2016}, but conventional grid-current-feedback designs face a bandwidth--grid-robustness compromise \cite{xu2018}. Robust admittance shaping, passivity-based design, and full-state feedback can directly shape multivariable terminal dynamics, but require model-based coordination of admittance, tracking, damping, and control effort \cite{cobreces2018,xie2020,cecati2022}. Adaptive control instead relies on online grid-impedance estimation \cite{temiz2020}, while data-driven passivity synthesis requires frequency-response data \cite{madani2024}. Current-error compensation retains the current-control architecture and improves weak-grid operation \cite{givaki2019}, yet does not introduce an explicit independent output-admittance degree of freedom. These limitations motivate a local feedback mechanism that reshapes output admittance without an external-network model while retaining the original current-loop dynamics.

This paper adapts the decoupled model-free control (MFC) architecture in \cite{li2026dmfc} to a three-phase current-controlled inverter for output-admittance reshaping. The added controller uses no external-grid or neighboring-converter model. A synchronous-frame Norton formulation derives a tracking-preserving condition under which the added local degree of freedom reshapes the output admittance while leaving the nominal current-tracking matrix unchanged. The generalized Nyquist criterion (GNC), applied to the complete $dq$ impedance matrices, and time-domain simulations assess the resulting interaction stability.

\section{Grid-Connected Inverter Model and Coupled Closed-Loop Characteristics}
\subsection{System Configuration and Current Controller}
\begin{figure}[t]
  \centering
  \includegraphics[width=0.95\columnwidth]{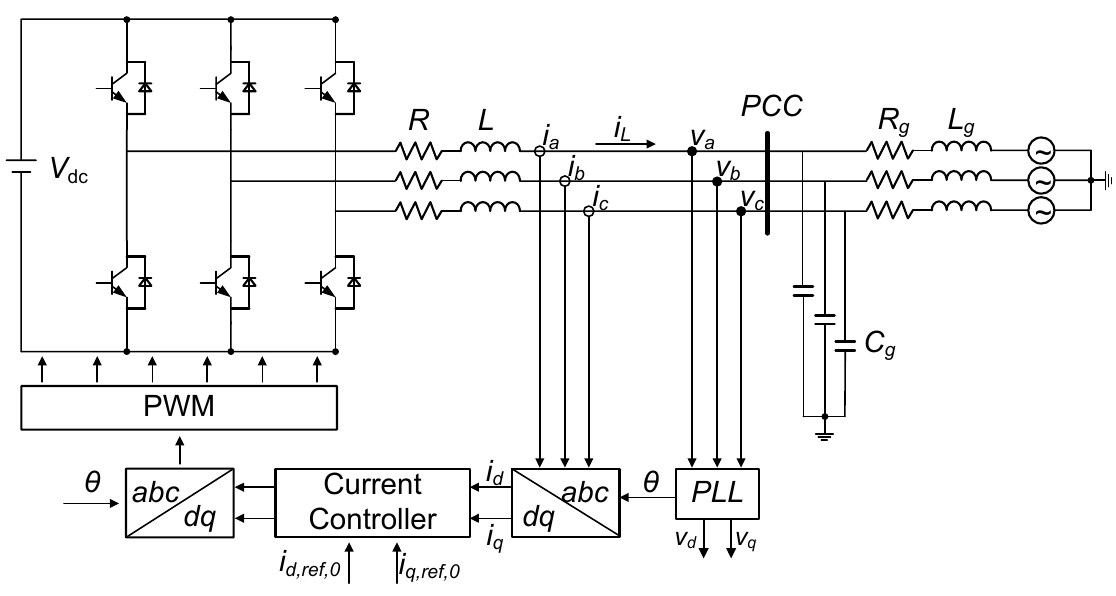}\\[-1pt]
  {\footnotesize (a)}\\[3pt]
  \includegraphics[width=\columnwidth]{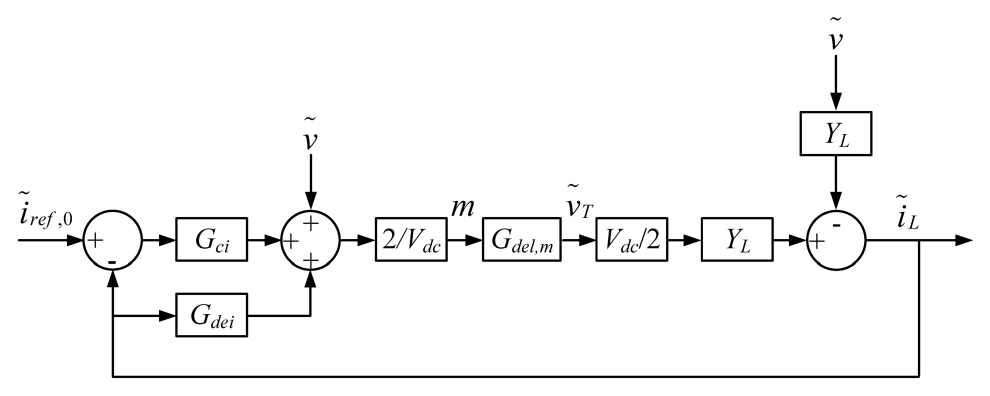}\\[-1pt]
  {\footnotesize (b)}
  \caption{Representative GFL inverter considered in this study: (a) system configuration with PLL-based synchronization; (b) small-signal current-control model with PLL dynamics neglected.}
  \label{fig:system}
\end{figure}

As a representative case study, Fig.~\ref{fig:system}(a) shows a grid-following (GFL) inverter with phase-locked-loop (PLL) synchronization and synchronous-frame current control. The dc link supplies a three-phase voltage-source inverter connected to the point of common coupling (PCC) through a filter inductance $L$ and resistance $R$. The external network comprises an equivalent shunt capacitance $C_g$ at the PCC, the series grid resistance $R_g$ and inductance $L_g$, and the grid voltage source. Throughout the paper, a tilde denotes a small-signal perturbation about the operating point; bold uppercase and lowercase quantities denote $2\times2$ matrices and two-component vectors, respectively. The external network is represented by the equivalent grid-impedance matrix $\mat Z_g$ seen from the PCC and the grid source-voltage perturbation $\widetilde{\mat v}_g$.

The analytical model describes small perturbations around a balanced three-phase steady state and neglects dc-link-voltage and PLL dynamics. All small-signal quantities are expressed in a common synchronous $dq$ frame at angular frequency $\omega_0$. The vectors $\widetilde{\mat i}_L$ and $\widetilde{\mat i}_{\mathrm{ref},0}$ denote the $dq$ filter-current perturbation and current-controller reference perturbation, respectively.

With the synchronous-frame cross-coupling matrix defined as $\mat J=\left[\begin{smallmatrix}0&-1\\1&0\end{smallmatrix}\right]$, the filter model is
\begin{equation}
 \widetilde{\mat i}_L=\mat Y_L
       (\widetilde{\mat v}_T-\widetilde{\mat v}),
 \label{eq:plant}
\end{equation}
where $\widetilde{\mat v}$ and $\widetilde{\mat v}_T$ are the PCC- and converter-side voltage perturbations, respectively, and the filter admittance matrix is
\begin{equation}
 \mat Y_L=\big[(Ls+R)\Id+\omega_0L\mat J\big]^{-1}.
 \label{eq:filter_admittance}
\end{equation}

Fig.~\ref{fig:system}(b) shows the small-signal current-control model including the proportional-integral (PI) current controller, inductor decoupling, unity PCC-voltage feedforward, and total digital delay. With the pulsewidth modulation (PWM) and dc-link scaling gains canceled, the converter-side voltage perturbation satisfies
\begin{equation}
 \widetilde{\mat v}_T
 =\Gdel\big[\mat G_{ci}\widetilde{\mat i}_{\mathrm{ref},0}
    +(\mat G_{dei}-\mat G_{ci})\widetilde{\mat i}_L
    +\widetilde{\mat v}\big],
\label{eq:controller}
\end{equation}
where the PI current-controller and inductor-decoupling matrices are, respectively,
\begin{align}
 \mat G_{ci}&=(k_{pi}+k_{ii}/s)\Id,
 \label{eq:current_controller}\\
 \mat G_{dei}&=\omega_0L\mat J .
 \label{eq:decoupling}
\end{align}
Here, $k_{pi}$ and $k_{ii}$ are the proportional and integral gains. The scalar transfer function $G_{\mathrm{del},m}$ models the total digital delay $T_d$, and its $dq$ matrix is $\Gdel=G_{\mathrm{del},m}\Id$. The analytical model uses the first-order Pad\'e approximation $G_{\mathrm{del},m}=(1-sT_d/2)/(1+sT_d/2)$.

The desired port behavior is that of a controlled current source: the injected current should follow its reference with limited sensitivity to PCC-voltage disturbances. This motivates a Norton representation, which exposes the reference-driven current and the voltage-dependent current separately.

\subsection{Norton Representation and Coupled Tracking--Admittance Dynamics}
Substituting \eqref{eq:controller} into \eqref{eq:plant}, collecting the current-dependent terms, and separating the reference- and voltage-driven components gives the Norton form
\begin{equation}
 \widetilde{\mat i}_L=\Gi\widetilde{\mat i}_{\mathrm{ref},0}
                   -\Ys\widetilde{\mat v},
 \label{eq:norton}
\end{equation}
where the nominal current-tracking matrix and output-admittance matrix are, respectively,
\begin{align}
 \Gi&=\big[\Id+\mat Y_L\Gdel(\mat G_{ci}-\mat G_{dei})\big]^{-1}
       \mat Y_L\Gdel\mat G_{ci},
 \label{eq:nominal_tracking}\\
 \Ys&=\big[\Id+\mat Y_L\Gdel(\mat G_{ci}-\mat G_{dei})\big]^{-1}
       \mat Y_L(\Id-\Gdel).
 \label{eq:output_admittance}
\end{align}

\begin{figure}[t]
 \centering
 \includegraphics[width=0.54\columnwidth]{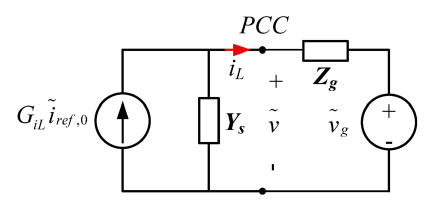}
 \caption{Norton equivalent of the inverter--grid interconnection.}
 \label{fig:norton}
\end{figure}

\begin{figure}[t]
 \centering
 \includegraphics[width=0.65\columnwidth]{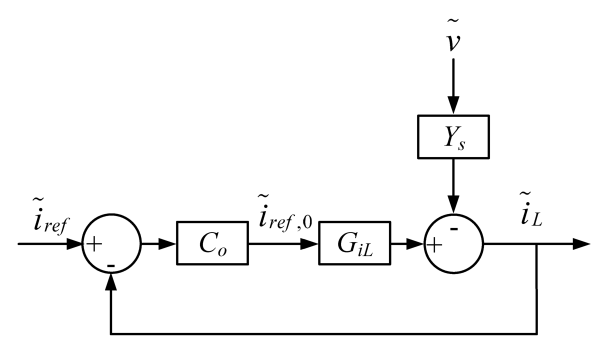}
 \caption{MFC feedback structure.}
 \label{fig:mfc_structure}
\end{figure}

Fig.~\ref{fig:norton} illustrates the associated port interconnection with PLL dynamics neglected. All transfer blocks in Figs.~\ref{fig:norton} and~\ref{fig:mfc_structure} represent $dq$ matrices. The matrix $\Gi$ describes nominal current tracking with $\widetilde{\mat v}=0$, whereas $\Ys$ describes voltage-to-current sensitivity with the reference perturbation set to zero. An ideal controlled current source has $\Ys=\mat 0$. Reducing this disturbance channel over the relevant frequencies therefore improves the inverter's current-source behavior.

The common inverse factor in \eqref{eq:nominal_tracking} and \eqref{eq:output_admittance} couples tracking and admittance: retuning $\mat G_{ci}$ or $\mat G_{dei}$ changes both. Conventional active damping exhibits a bandwidth--robustness compromise \cite{xu2018}, while robust admittance shaping requires model-based coordination of harmonic attenuation and resonance damping \cite{cobreces2018}. This motivates an additional control degree of freedom that reshapes admittance while preserving tracking without an external-network model.

\section{Model-Free Control for Output-Admittance Reshaping}
\subsection{Control Structure and Admittance Shaping}
The added MFC transfer matrix $\Co$ in Fig.~\ref{fig:mfc_structure} converts the error between the external reference $\widetilde{\mat i}_{\mathrm{ref}}$ and the measured current into the current-controller input $\widetilde{\mat i}_{\mathrm{ref},0}$:
\begin{equation}
 \widetilde{\mat i}_{\mathrm{ref},0}
   =\Co(\widetilde{\mat i}_{\mathrm{ref}}-\widetilde{\mat i}_L).
 \label{eq:mfc_law}
\end{equation}

The existing current controller, decoupling, and feedforward remain within the local subsystem represented by \eqref{eq:norton}. Substituting the MFC feedback law in \eqref{eq:mfc_law} into the nominal Norton model in \eqref{eq:norton} yields the MFC closed-loop port relation
\begin{equation}
 \widetilde{\mat i}_L
   =\Gim\widetilde{\mat i}_{\mathrm{ref}}-\Ym\widetilde{\mat v},
 \label{eq:mfc_port}
\end{equation}
where $\Gim$ and $\Ym$ are the MFC closed-loop current-tracking and output-admittance matrices, respectively,
\begin{equation}
 \Gim=(\Id+\Gi\Co)^{-1}\Gi\Co,
 \label{eq:mfc_tracking}
\end{equation}
\begin{equation}
 \Ym=(\Id+\Gi\Co)^{-1}\Ys.
 \label{eq:mfc_admittance}
\end{equation}

The MFC controller $\Co$ forms an outer current-feedback loop and provides an additional degree of freedom. When $\Co$ is selected to approximately preserve the nominal reference-to-current dynamics, i.e., $\Gim\approx\Gi$, \eqref{eq:mfc_tracking} and \eqref{eq:mfc_admittance} give the approximate admittance-shaping relation
\begin{equation}
 \Ym\approx\Co^{-1}\Ys.
 \label{eq:mfc_tracking_preserving_admittance}
\end{equation}

Equation~\eqref{eq:mfc_tracking_preserving_admittance} shows that suitable $\Co$ attenuates output admittance while closely preserving tracking. Reduced PCC-voltage sensitivity approaches ideal controlled-current-source behavior, weakening inverter--network interaction and improving interconnection stability. The shaping relations contain only local converter quantities and require no external-network model.

\subsection{Generalized Nyquist Stability Assessment}
With the grid-voltage perturbation suppressed, define the series grid-branch impedance as $\mat Z_{RL,g}(s)=(R_g+sL_g)\Id+\omega_0L_g\mat J$ and the PCC-capacitor admittance as $\mat Y_{C,g}(s)=C_g(s\Id+\omega_0\mat J)$. The equivalent external-network impedance seen from the PCC is
\begin{equation}
 \mat Z_g(s)=\big[\mat Z_{RL,g}^{-1}(s)+\mat Y_{C,g}(s)\big]^{-1}.
 \label{eq:grid_driving_impedance}
\end{equation}

Following the impedance-based stability criterion in \cite{sun2011,wen2017}, the return-ratio matrix of the inverter--grid interconnection in Fig.~\ref{fig:norton} without MFC is
\begin{equation}
 \mat L_0(s)=\mat Z_g(s)\Ys(s),
 \label{eq:return_no_mfc}
\end{equation}
and that with MFC is
\begin{equation}
 \mat L_{\mathrm{MFC}}(s)=\mat Z_g(s)\Ym(s).
 \label{eq:return_mfc}
\end{equation}

For the case-study GNC calculations, the same return-ratio construction uses inverter admittance matrices including PLL dynamics in place of the simplified expressions derived above. Because the converter and grid subsystems are individually stable in all cases considered, the GNC classifies the interconnection as stable when the eigenloci of the applicable return-ratio matrix do not encircle $-1$ \cite{wen2017}.

\section{Case Study and Stability Validation}
The case study uses a three-phase averaged inverter model in all results in Figs.~\ref{fig:tracking_steps}--\ref{fig:iq_boundary}. Within each comparison, the cases without and with MFC use the same power circuit, current controller, delay, grid, operating point, and event. Table~\ref{tab:parameters} lists the common system and control settings together with the nominal grid parameters. Alternative grid conditions are stated in the corresponding subsections.

\begin{table}[t]
\caption{Common System, Control, and Nominal Grid Parameters}
\label{tab:parameters}
\centering
\footnotesize
\setlength{\tabcolsep}{3pt}
\begin{tabular}{lll}
\toprule
Parameter & Symbol & Value\\
\midrule
DC-link voltage & $V_{dc}$ & $400~\mathrm{V}$\\
Nominal PCC voltage & $(v_d,v_q)$ & $(100,\,0)~\mathrm{V}$\\
Fundamental frequency & $f_0$ & $50~\mathrm{Hz}$\\
Switching frequency & $f_{sw}$ & $20~\mathrm{kHz}$\\
Filter inductance & $L$ & $2~\mathrm{mH}$\\
Filter resistance & $R$ & $50~\mathrm{m}\Omega$\\
Nominal grid resistance & $R_g$ & $0.2108~\Omega$\\
Nominal grid inductance & $L_g$ & $0.671~\mathrm{mH}$\\
Equivalent grid shunt capacitance & $C_g$ & $410~\mu\mathrm{F}$\\
Current-controller gains & $k_{pi},k_{ii}$ & $8,\ 200$\\
Control delay & $T_d$ & $1.5/f_{sw}$\\
Current-loop time constant & $\tau$ & $5/f_{sw}$\\
Nominal current reference & $(i_{d,\mathrm{ref}},i_{q,\mathrm{ref}})$ & $(20,\,10)~\mathrm{A}$\\
\bottomrule
\end{tabular}
\end{table}

As a concrete example for this case study, we select an outer controller to closely preserve the nominal current-tracking dynamics, i.e., $\Gim(s)\approx\Gi(s)$ over the current-control bandwidth. With the parameters in Table~\ref{tab:parameters} and the first-order Pad\'e delay model, the $d$- and $q$-axis outer loops in Fig.~\ref{fig:mfc_structure} use
\begin{equation}
 C_{o,d}(s)=C_{o,q}(s)
 =1+\frac{4000-0.15s}{s+3.75\times10^{-5}s^2}.
 \label{eq:case_co}
\end{equation}
This transfer function is applied separately to the $d$- and $q$-axis current errors; no external-grid parameter is required.

\subsection{Reference-Tracking Preservation}
\begin{figure}[!ht]
 \centering
 \includegraphics[width=0.82\columnwidth,height=0.26\columnwidth,keepaspectratio]{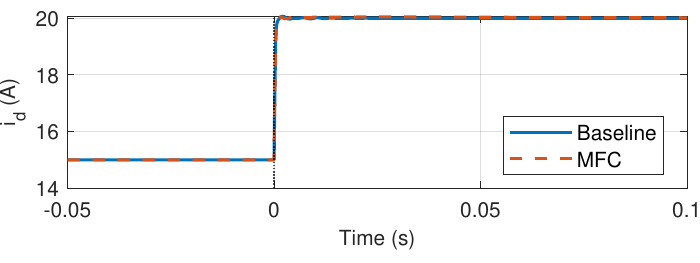}\\[-3pt]
 {\footnotesize (a)}\\[1pt]
 \includegraphics[width=0.82\columnwidth,height=0.26\columnwidth,keepaspectratio]{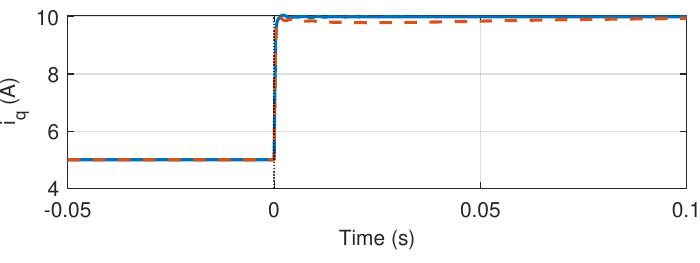}\\[-3pt]
 {\footnotesize (b)}
 \caption{Current-reference step responses under the nominal grid condition ($\mathrm{SCR}=10$, $X/R=1$): (a) $d$ axis; (b) $q$ axis.}
\label{fig:tracking_steps}
\end{figure}

The reference-tracking tests use the nominal grid parameters in Table~\ref{tab:parameters}, corresponding to a short-circuit ratio $\mathrm{SCR}=10$ and a grid reactance-to-resistance ratio $X/R=1$. Separate 5-A reference steps are applied. In Fig.~\ref{fig:tracking_steps}, the maximum absolute differences between the two responses without and with MFC, normalized by the 5-A step amplitude, are 1.4\% and 4.2\% for the $d$ and $q$ axes, respectively. These results indicate that the added MFC loop closely preserves the nominal reference-to-current dynamics.

\subsection{Output-Admittance Reduction}
\begin{figure}[!ht]
 \centering
 \includegraphics[width=0.90\columnwidth,height=0.48\columnwidth,keepaspectratio]{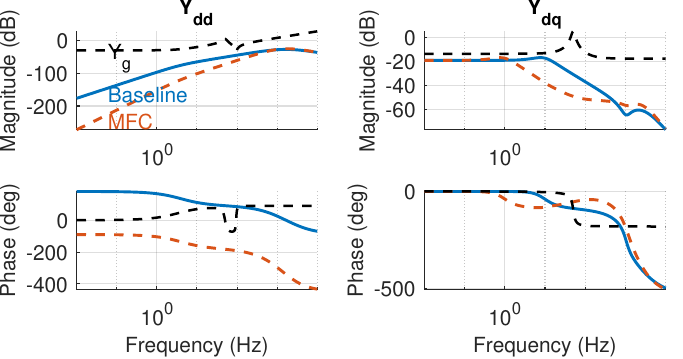}\\[1pt]
 \includegraphics[width=0.90\columnwidth,height=0.48\columnwidth,keepaspectratio]{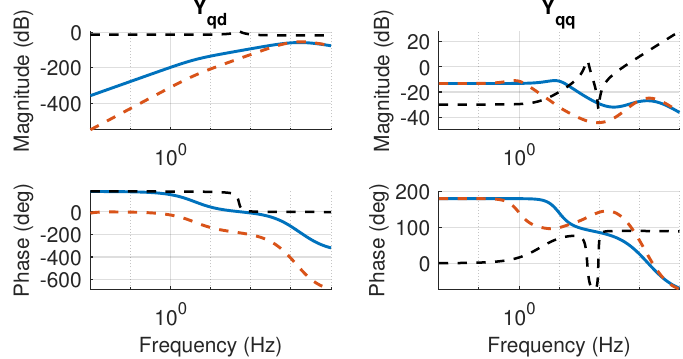}
 \caption{Real-$dq$ admittance responses $Y_{dd}$, $Y_{dq}$, $Y_{qd}$, and $Y_{qq}$ at $\mathrm{SCR}=1$ and $X/R=10.5$.}
\label{fig:admittance}
\end{figure}

The output-admittance comparison uses $\mathrm{SCR}=1$ and $X/R=10.5$. Fig.~\ref{fig:admittance} compares the four entries $Y_{dd}$, $Y_{dq}$, $Y_{qd}$, and $Y_{qq}$ of the output-admittance matrix in the $dq$ frame. It shows that MFC reduces the magnitude of every entry by more than 65\% from 10 to 200~Hz, corresponding to an attenuation of more than 9.1~dB. According to \eqref{eq:case_co}, $\Co\to\Id$ at high frequencies; for the model with PLL dynamics neglected, \eqref{eq:mfc_tracking_preserving_admittance} gives $\Ym\to\Ys$. These results show reduced terminal-voltage sensitivity, consistent with the mechanism described by \eqref{eq:mfc_tracking_preserving_admittance}.

\subsection{Stability-Boundary Validation}
GNC is used to map stability boundaries under grid-impedance and current-operating-point variations; representative predicted crossings are then verified in averaged-model simulations.

\subsubsection{Grid-Impedance Boundary}
\begin{figure}[!ht]
 \centering
 \includegraphics[width=0.84\columnwidth]{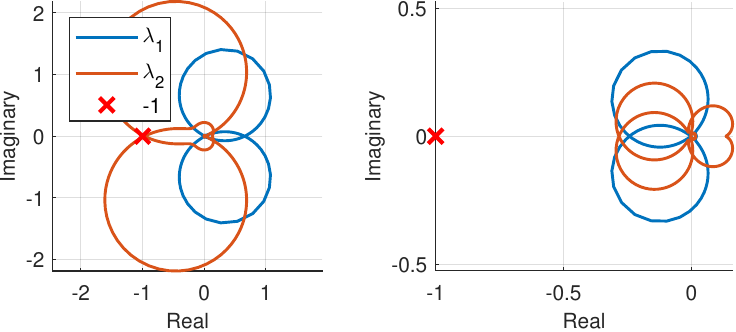}\\[-3pt]
 \parbox{0.48\columnwidth}{\centering\footnotesize (a)}\hfill
 \parbox{0.48\columnwidth}{\centering\footnotesize (b)}
 \caption{GNC eigenloci before the SCR step ($\mathrm{SCR}=2$, $X/R=10.5$): (a) without MFC; (b) with MFC.}
 \label{fig:nyquist_scr2}
\end{figure}

\begin{figure}[!ht]
 \centering
 \includegraphics[width=0.84\columnwidth]{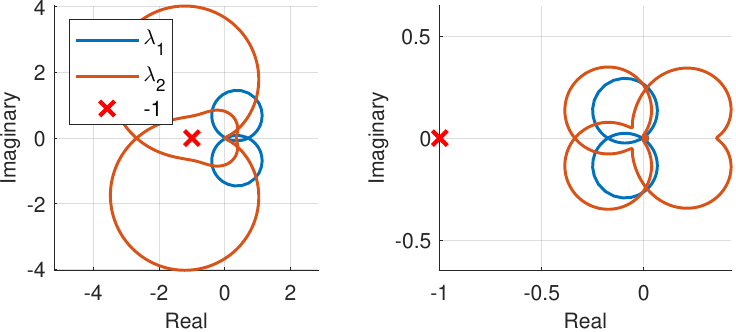}\\[-3pt]
 \parbox{0.48\columnwidth}{\centering\footnotesize (a)}\hfill
 \parbox{0.48\columnwidth}{\centering\footnotesize (b)}
\caption{GNC eigenloci after the grid-strength reduction ($\mathrm{SCR}=1$, $X/R=10.5$): (a) without MFC; (b) with MFC.}
\label{fig:nyquist_scr1}
\end{figure}

At $X/R=10.5$, Figs.~\ref{fig:nyquist_scr2} and~\ref{fig:nyquist_scr1} show the GNC eigenloci at $\mathrm{SCR}=2$ and $\mathrm{SCR}=1$, respectively. As SCR decreases from 2 to 1, the baseline locus changes from no encirclement to an encirclement of the critical point $(-1, j0)$, predicting loss of stability. With MFC, the loci contract toward the origin without encircling the critical point $(-1, j0)$, predicting stability under both grid conditions.

\begin{figure}[!ht]
 \centering
 \begin{minipage}[b]{0.485\columnwidth}
  \centering
 \includegraphics[width=\linewidth,height=0.88\columnwidth,keepaspectratio]{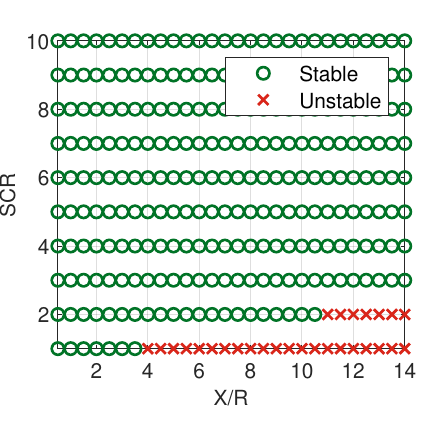}\\[-3pt]
  {\footnotesize (a)}
 \end{minipage}\hfill
 \begin{minipage}[b]{0.485\columnwidth}
  \centering
 \includegraphics[width=\linewidth,height=0.88\columnwidth,keepaspectratio]{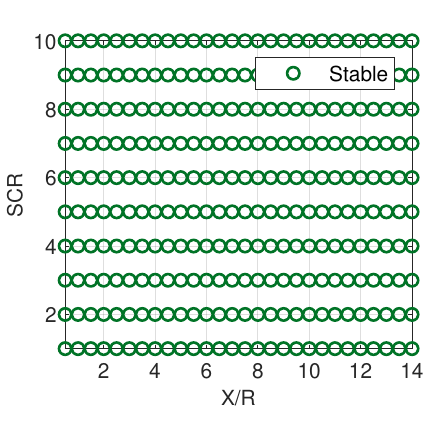}\\[-3pt]
  {\footnotesize (b)}
 \end{minipage}
 \caption{GNC-predicted SCR--$X/R$ stability regions: (a) without MFC; (b) with MFC.}
\label{fig:scr_map}
\end{figure}

Applying this test over $1\leq\mathrm{SCR}\leq10$ and $0.5\leq X/R\leq14$ yields Fig.~\ref{fig:scr_map}. MFC increases the GNC-stable coverage from 90.0\% to 100\%. These results show that MFC extends the admissible grid-impedance range to lower SCR and higher $X/R$, i.e., weaker and more inductive grids.

\begin{figure}[!htb]
 \centering
 \includegraphics[width=0.65\columnwidth]{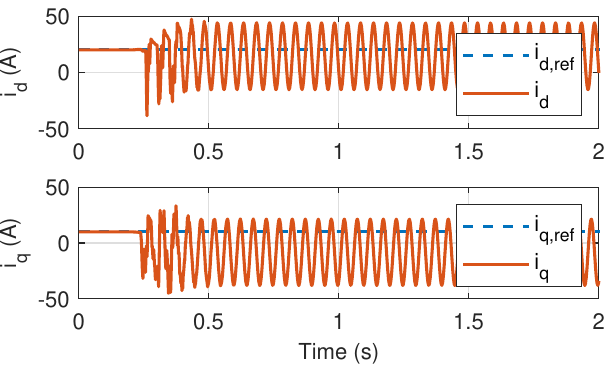}\\[-3pt]
 {\footnotesize (a)}\\[1pt]
 \includegraphics[width=0.65\columnwidth]{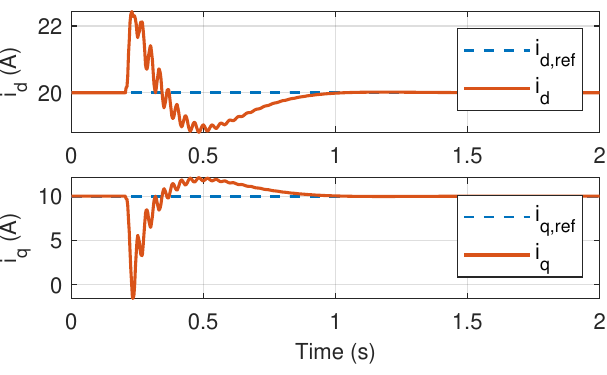}\\[-3pt]
 {\footnotesize (b)}
 \caption{Time-domain validation for an SCR step from 2 to 1 at $t=0.2~\mathrm{s}$ and $X/R=10.5$: (a) without MFC; (b) with MFC.}
\label{fig:grid_transition}
\end{figure}

Fig.~\ref{fig:grid_transition} tests the predicted SCR boundary crossing in the averaged model. The response without MFC develops a sustained coupled oscillation, whereas MFC damps the oscillation and restores both currents to their references. These responses demonstrate improved interconnection stability with MFC under the weakened-grid condition and corroborate the GNC prediction.

\subsubsection{Operating-Point Boundary}
\begin{figure}[!ht]
 \centering
 \begin{minipage}[b]{0.485\columnwidth}
  \centering
 \includegraphics[width=\linewidth,height=0.88\columnwidth,keepaspectratio]{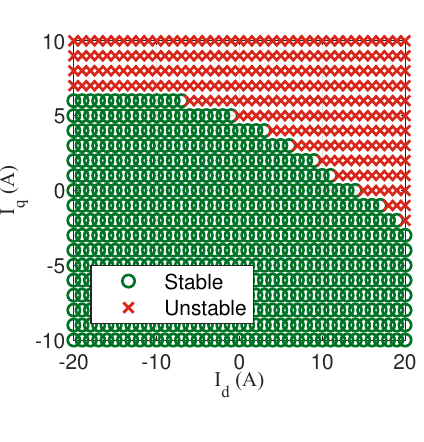}\\[-3pt]
  {\footnotesize (a)}
 \end{minipage}\hfill
 \begin{minipage}[b]{0.485\columnwidth}
  \centering
 \includegraphics[width=\linewidth,height=0.88\columnwidth,keepaspectratio]{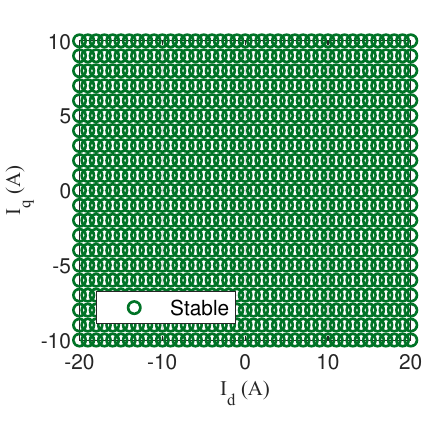}\\[-3pt]
  {\footnotesize (b)}
 \end{minipage}
 \caption{GNC-predicted current-operating-point stability regions at $\mathrm{SCR}=1.2$ and $X/R=14$: (a) without MFC; (b) with MFC.}
\label{fig:idq_map}
\end{figure}

At $\mathrm{SCR}=1.2$ and $X/R=14$, applying GNC over the sampled current-reference pairs yields the operating-point map in Fig.~\ref{fig:idq_map}. MFC increases the predicted stable coverage from 68.3\% to 100\%. These results show that MFC extends the admissible range of $dq$-current references under the same weak-grid condition.

\begin{figure}[!ht]
 \centering
 \includegraphics[width=0.65\columnwidth]{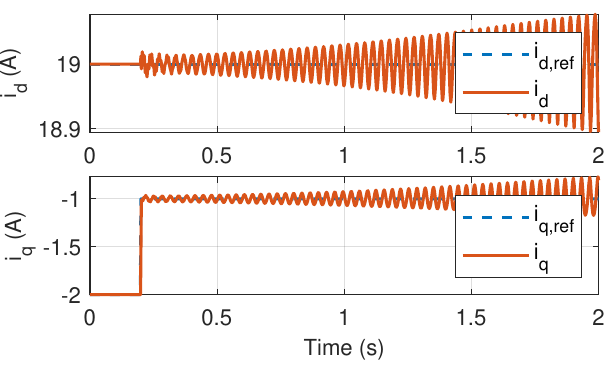}\\[-3pt]
 {\footnotesize (a)}\\[1pt]
 \includegraphics[width=0.65\columnwidth]{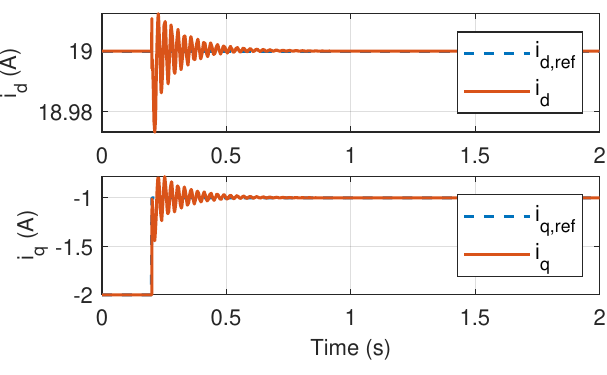}\\[-3pt]
 {\footnotesize (b)}
 \caption{Time-domain validation for an operating-point boundary crossing at $i_{d,\mathrm{ref}}=19~\mathrm{A}$, $\mathrm{SCR}=1.2$, and $X/R=14$: (a) without MFC; (b) with MFC.}
\label{fig:iq_boundary}
\end{figure}

The simulations in Fig.~\ref{fig:iq_boundary} validate a representative crossing of the baseline boundary predicted in Fig.~\ref{fig:idq_map}(a). At $i_{d,\mathrm{ref}}=19~\mathrm{A}$, the $i_{q,\mathrm{ref}}$ step from $-2$ to $-1~\mathrm{A}$ produces growing oscillations without MFC but decaying oscillations with MFC. These responses demonstrate improved stability with MFC at the tested operating point and corroborate the GNC classifications.

\section{Conclusion}
This paper has developed MFC-based output-admittance reshaping for a current-controlled grid-connected inverter. The outer-loop feedback controller $\Co$ of MFC reshapes the output admittance while closely preserving the nominal reference-to-current dynamics without requiring an external-grid model. The results show lower $dq$-frame output-admittance magnitudes, closely matched tracking, and enlarged stability regions under grid-impedance and current-reference variations. Time-domain responses corroborate both stability assessments.

\bibliographystyle{IEEEtran}
\bibliography{references}
\end{document}